\documentclass[twocolumn,showpacs,preprintnumbers,amsmath,amssymb]{revtex4}

\usepackage{graphicx}
\usepackage{dcolumn}
\usepackage{bm}
\usepackage{epsfig}
\usepackage{epstopdf}

\begin{document}


\title{Parametric downconversion with optimized spectral properties in nonlinear photonic crystals}

\author{Mar{\'i}a Corona and Alfred B. U'Ren}
\affiliation{Centro de Investigaci\'{o}n Cient\'{i}fica y de
Educaci\'{o}n Superior de Ensenada (CICESE), Baja California,
22860, Mexico}

\date{\today}

%
\newcommand{\epsfg}[2]{\centerline{\scalebox{#2}{\epsfbox{#1}}}}

\begin{abstract}
We study the joint spectral properties of photon pairs generated
by spontaneous parametric down-conversion in a one-dimensional
nonlinear photonic crystal in a collinear, degenerate, type-II
geometry.  We show that the photonic crystal properties may be
exploited to compensate for material dispersion and obtain photon
pairs that are nearly factorable, in principle, for arbitrary
materials and spectral regions, limited by the ability to
fabricate the nonlinear crystal with the required periodic
variation in the refractive indices for the ordinary and
extraordinary waves.
\end{abstract}

\maketitle



\section{Introduction}

Two-photon states with specific continuous variable-entanglement
properties are required for a number of quantum-information-processing applications. In particular, pure-state single photons,
crucial for applications relying on interference between
independently prepared single photons, such as quantum computing
with linear optics~\cite{kok05}, entanglement
swapping~\cite{pan98}, and teleportation~\cite{bouwmeester97}, can
be heralded only from \textit{factorable} photon pairs, where no
correlations exist between the constituent single
photons~\cite{uren05}.    Let us note that factorability may be
imposed by postselecting an appropriate subensemble of photon
pairs through spectral and/or spatial filtering; this can be done
only at the cost of a typically drastic reduction in count rate
and source brightness.  Scalable quantum-information processing
requires the elimination of this type of postselection. It has
furthermore been shown that, for quantum-information applications
involving multiple-pair generation resulting from high-gain
parametric down-conversion, spectral or spatial filtering is likewise
ineffectual.~\cite{rohde06}. In addition, the coexistence of
factorability and a large generated bandwidth is important for
some applications, such as the generation of ultrashort
Fourier-transform-limited heralded single photons~\cite{uren07}.
The process of spontaneous parametric down-conversion represents a
well-established method for the generation of photon pairs,
leading to unparalleled flexibility in the resulting continuous-variable entanglement properties.  Indeed, the nonlinear crystal
dispersion characteristics in conjunction with temporal and
spatial structure in the pump beam may be exploited to engineer
the type and degree of correlations present in spectral and
transverse wave-vector degrees of freedom \textit{at the source},
thus eliminating the need for photon pair filtering.

The use of a broadband pump is essential in the context of the
generation of photon pairs with spectrally engineered properties;
indeed, a monochromatic pump can access only a one-dimensional
subspace of signal-idler frequency space, which precludes certain
states of interest. Previous work has established the central role
that is played by group velocity mismatch terms between the three fields
involved in parametric down-conversion (pump, signal, and idler)
in the determination of the resulting photon pair
properties. Thus, in Ref.~\cite{grice01}, it was shown that if the
pump pulse propagates at a higher group velocity than one of the
generated photons but lower than the conjugate generated photon,
then it becomes possible to emit factorable photon pairs.   An
important limitation of this technique is that the group velocity
condition which must be satisfied occurs only for specific
materials, at specific spectral
ranges~\cite{giovannetti02,kuzucu05}, typically in the infrared
where single-photon detection is unfortunately not well developed.
A number of techniques have been proposed and in some cases
implemented which enable effective control over the photon pair
entanglement properties in the spectral domain unconstrained by
material dispersion.  In one such technique, the effective group
velocities experienced by the three fields involved are modified
by a pair of diffraction gratings placed before and after the
nonlinear crystal so as to generate photon pairs with tunable
spectral characteristics~\cite{torres05,hendrych06}.
Alternatively, the spectral content of two photon states may be
restricted to the modes supported by a nonlinear cavity which
leads, for a short, high-finesse cavity, to factorable, narrowband
photon pair generation~\cite{raymer05}.  Likewise, on the one hand
a transversely pumped source where signal and idler photons are
emitted in a counterpropagating waveguided
geometry~\cite{walton04} and on the other hand noncollinear parametric down-conversion (PDC) where a specific relationship between transverse and longitudinal
phase matching is attained~\cite{uren03} can both lead to states
with spectrally engineered properties.  Another route is the use
of crystal superlattices, where the dispersion in short crystal
segments is compensated by birrefringent compensators, permitting
two-photon states with a wide range of possible spectral
entanglement characteristics~\cite{uren06}.

Previous theoretical work has explored the use of non linear
photonic crystals in the context of the process of spontaneous
parametric down-conversion.   In particular,
Refs.~\cite{dood04,irvine05} study the potential of
semiconductor-based nonlinear one-dimensional photonic crystals to
yield phase-matching properties appropriate for the generation of
polarization-entangled photon pairs.  Likewise, it has been shown
that even a short one-dimensional photonic crystal is capable of
generating a considerable photon pair flux due to field
localization in such structures~\cite{centini05, perina06,
vamivakas04}. In this paper we study the potential of exploiting
the properties of one-dimensional nonlinear photonic crystals in
order to generate photon pairs with engineered spectral
entanglement properties.  We concentrate on type-II, frequency
degenerate, collinear PDC.  Collinear operation is important
because it permits PDC in a waveguided geometry, which leads to
larger generation rates as well as to effective control over the
transverse spatial structure of the emitted light, crucial for
efficient single-mode fiber coupling and for spatial mode matching
in interference experiments~\cite{uren04}. Type-II operation for
which the signal and idler photons are orthogonally polarized is
important because it enables efficient photon pair splitting. We
show that a weak index of refraction modulation present in an
otherwise standard birefringent nonlinear crystal can be exploited
to attain the group velocity conditions, in addition to basic
phase matching, required for factorable photon pair generation.

\section{One-dimensional nonlinear photonic crystals}

In this paper we analyze the generation of photon pairs by the
process of parametric down-conversion in a $\chi{(2)}$ material
characterized by a spatial periodicity in its linear optical
properties. In particular, we study PDC produced by a nonlinear photonic crystal (NLPC) based on a standard bulk nonlinear
crystal with uniaxial birefringence which has been modified from its
natural state by an appropriate physical mechanism so as to yield
a periodic variation of the index of refraction for both
polarizations, while maintaining the nonlinearity constant
throughout the crystal thickness. Concretely, within one period of
the resulting Bragg grating, we assume that the index of refraction
for the ordinary ($\mu=o$) and the extraordinary ray ($\mu=e$) are given by

\begin{equation}\label{Ec:index}
 n_\mu(\omega,z)=\left\{ \begin{array}{ll}
n_{\mu 1}(\omega),\ \ \ 0<z<a, \\
n_{\mu 2}(\omega),\ \ \ a<z<\Lambda.
\end{array} \right.
\end{equation}

\begin{figure}[ht]
\includegraphics[width=3in]{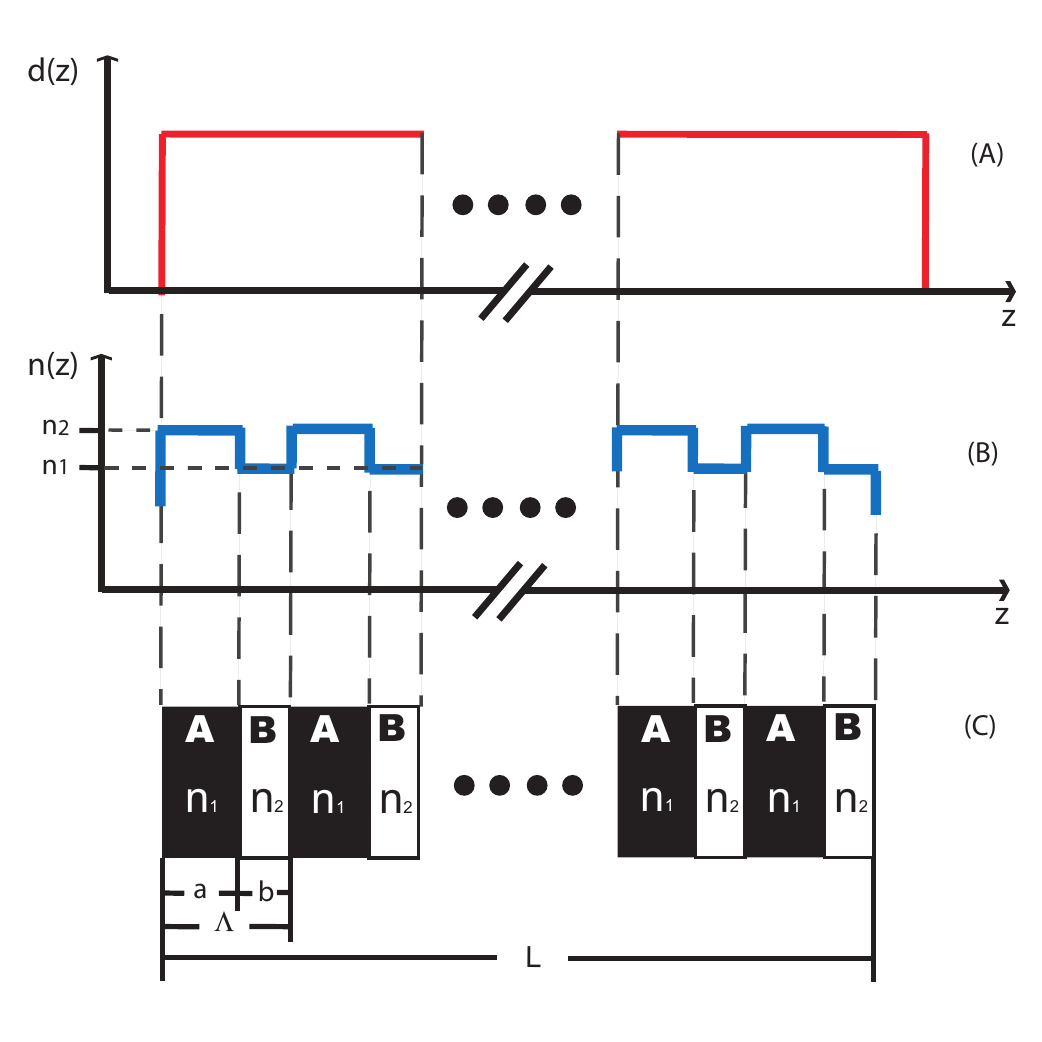} \caption{(color online)
Schematic of a one-dimensional, nonlinear photonic crystal.
(a)Second order nonlinearity.  (b) Refractive index, shown for one
of the polarizations.  (c) Representation of the periodic material,
with $b=\Lambda-a$. \label{Fig:Bragg}}
\end{figure}

This unit cell is assumed to be replicated throughout the crystal
length. Here, $\Lambda$ represents the Bragg period and
$a/\Lambda$ is the duty cycle.   We refer to such a material as a
one-dimensional nonlinear photonic crystal. Figure~\ref{Fig:Bragg}
shows a crystal schematic where $A$ indicates zones characterized
by indices of refraction $n_{o1}$ and $n_{e1}$ and $B$ indicates
zones characterized by indices of refraction $n_{o2}$ and
$n_{e2}$. We assume that in zones $A$ the crystal has been left in
its natural state, while in zones $B$ it has undergone index
change. For simplicity, we assume that zones $A$ and $B$ are of
equal length (leading to an $a/\Lambda=0.5$ duty cycle) and we
likewise assume that the permittivity contrast $\alpha$,
equivalent to the index square contrast, is frequency independent,

\begin{equation}\label{Ec:permittivity_contrast}
\alpha=\frac{2[\epsilon_{\mu1}(\omega)-\epsilon_{\mu2}(\omega)]}{\epsilon_{\mu1}(\omega)+\epsilon_{\mu2}(\omega)}
=\frac{n_{\mu 1}(\omega)^2-n_{\mu
2}(\omega)^2}{\overline{n}_\mu(\omega)^2},
\end{equation}
where

\begin{equation}\label{Ec:index_prom}
\overline{n}_\mu(\omega)^2=[ n_{\mu 1}(\omega)^2+n_{\mu
2}(\omega)^2]/2.
\end{equation}
with $\mu=e,o$. Throughout this paper we will assume that propagation of the three
fields involved is normal to each of the interfaces between zones
$A$ and $B$.  Under these conditions, such a material can exhibit,
for each of the polarizations, a so-called photonic band gap
centered at each of the Bragg wavelengths

\begin{equation}\label{Ec:lambda_Bragg}
\lambda^{Bragg,m}_{\mu}=2 \overline{n}_{\mu} \Lambda/m,
\end{equation}
where $m=1,2,3,...$, and where $\mu=e,o$.   Within each band gap, for a sufficient
crystal thickness, light is efficiently reflected, while for
frequencies outside the band gap, light propagates in the form of
so-called Bloch waves which can be written as

\begin{equation}\label{Ec:Bloch_waves}
E(z,t)=E_K(z) \exp\{i [K(\omega) z-\omega t]\}.
\end{equation}

Here, $E_K(z)$ is the Bloch envelope, which exhibits a spatial
periodicity matching that of the material, i.e.,
$E_K(z+\Lambda)=E_K(z)$, while $K(\omega)$ represents the Bloch
wave number.  Note that, for a continuous material without Bragg
grating, the Bloch envelope reduces to a constant, and therefore
the Bloch wave reduces to a plane wave.  Following a coupled-mode
analysis where the spatial periodicity in the permittivity is
assumed to be well represented by a weak perturbation to the
material permittivity, it can be shown that the Bloch wave number
in the vicinity of $K=m \pi/\Lambda$ can be expressed
as \cite{yariv84}

\begin{equation}\label{Ec:wavenumber}
K_\mu^{(m)}(\omega)=\pi m  /\Lambda \pm\sqrt{(\Delta
\beta_\mu^{(m)}/2)^2-|\kappa_\mu^{(m)}|^2}.
\end{equation}
Here, $\kappa_\mu^{(m)}$  represents the coupling coefficient between the
forward and backward waves

\begin{equation}\label{Ec:couple_coef}
\kappa_\mu^{(m)}=i [1-\cos(m \pi)]\alpha \overline{k}_\mu(\omega)/ (4 \pi m),
\end{equation}
and $\Delta \beta_\mu^{(m)}$ represents the Bragg phase mismatch between
these two waves,

\begin{equation}\label{Ec:Bmismatch}
\Delta \beta_\mu^{(m)}= 2 \overline{k}_\mu(\omega)-2\pi m/\Lambda,
\end{equation}
where $\overline{k}_\mu=\overline{n}_\mu \omega/c$ characterizes
the underlying material dispersion; note that
Eq.(\ref{Ec:lambda_Bragg}) follows from the condition $\Delta
\beta^{(m)}=0$.  A photonic band gap appears when $\kappa_\mu^{(m)}
\neq 0$ and its effects become appreciable for $\Delta
\beta_\mu^{(m)} \approx 0$.  The band gap boundaries $\omega_{min}$
and $\omega_{max}$ are obtained by solving for the frequencies
such that the argument of the square root in
Eq.(\ref{Ec:wavenumber}) vanishes. For a forward-propagating wave,
the minus sign in front of the square root in
Eq.(\ref{Ec:wavenumber}) should be used for $\omega<\omega_{min}$
while the plus sign should be used for $\omega>\omega_{max}$.  Let
us note that for a $0.5$ duty cycle, Eq.(\ref{Ec:couple_coef})
tells us that, even order band gaps ($m=2,4,6...$) are suppressed.

In what follows, we will concentrate on first order ($m=1$)
band gaps, though the analysis below could be employed for any
order. In particular, for practical implementations it may be
challenging to fabricate the required Bragg gratings with periods
in the region of hundreds of nanometers compatible with modified
dispersive characteristics in the visible; alternatively, it is
possible to use longer periods, together with higher-order
band gaps.  The dispersive properties of NPLCs are characterized
by the function $K(\omega)$, where from this point we omit the
order superscript. For propagating waves at frequencies in the
vicinity of the band gap boundaries, $K(\omega)$ can differ
substantially from the underlying material dispersion relation
$\overline{k}(\omega)$.  In particular, group velocities can be
strongly reduced, while group velocity dispersion terms (as well
as higher-order frequency derivatives of the wavenumber) can
increase markedly, from their values associated with material-only
dispersion.

\begin{figure}[ht]
\centering \epsfg{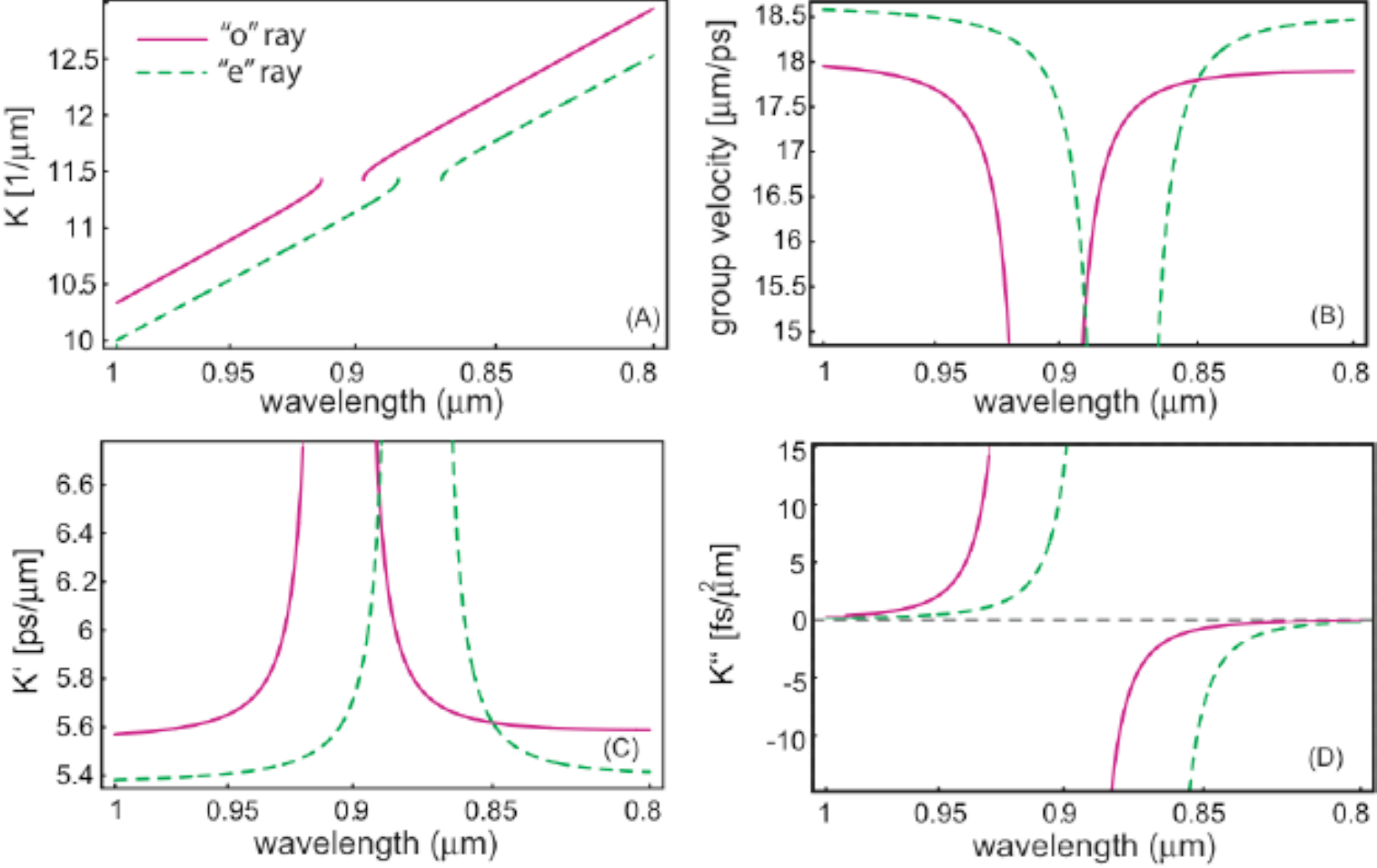}{0.5} \caption{(Color online)
Plots as a function of frequency of (a) the Bragg wave number
$K(\omega)$, (b) the group velocity $1/K'(\omega)$, (c)
$K'(\omega)$, and (d) group velocity dispersion coefficient
$K''(\omega)$; for comparison, the dotted horizontal line
indicates the magnitude of the GVD coefficient for SF-10, a
particularly dispersive glass. \label{Fig:dispersion}}
\end{figure}

In order to illustrate these effects, let us consider a specific
example of a one-dimensional NLPC, based on a $\beta$-barium-borate
(BBO) crystal. We assume that the Bragg period is given by
$\Lambda=279.1$ nm with a $a/\Lambda=0.5$ duty cycle, the angle
subtended by the pump and the optic axis (to be referred to as the
propagation angle) is $41.8^{\circ}$, and the permittivity
contrast is $\alpha=0.027$. Figure~\ref{Fig:dispersion} shows the
resulting dispersive properties.  Figure~\ref{Fig:dispersion} (a) shows, for each of the two
polarizations, a plot of the Bloch wave number $K(\omega)$, showing
clearly a band gap for each of the polarizations, within which
$K(\omega)$ becomes complex-valued. Figure~\ref{Fig:dispersion} (b)
shows a plot of the group velocities for each of the two
polarizations, exhibiting a marked reduction near the band gap
boundaries. Figure~\ref{Fig:dispersion} (c) shows the first
frequency derivative of $K(\omega)$, or the reciprocal group
velocity.  Finally, Figure~\ref{Fig:dispersion} (d) shows a plot of
the group velocity dispersion (GVD) term for each of the
polarizations.  It is apparent from the plot that GVD can be
greatly enhanced near the band gap boundaries, and likewise that it
is possible to obtain both positive and negative GVD.

\section{PDC in one-dimensional photonic nonlinear crystals}

Following a standard perturbative approach, the quantum state
describing photon pairs produced by parametric down-conversion in the
spontaneous limit may be expressed as

\begin{equation}\label{Ec:qstate}
|\psi(t)\rangle\approx
\left[1+\frac{1}{i\hbar}\int_0^{t}dt'\hat{H}(t')\right]|\textrm{vac}\rangle,
\end{equation}
where $|\mbox{vac}\rangle$ denotes the vacuum and $\hat{H}$ is the
interaction Hamiltonian

\begin{equation}\label{Ec:Hamiltonian}
\hat{H}(t)=\varepsilon_0\int_V d V\quad d(\vec{r})
\hat{E}_p^{(+)}(\vec{r},t)\hat{E}_s^{(-)}(\vec{r},t)\hat{E}_i^{(-)}(\vec{r},t)+\textrm{H.C.}
\end{equation}
Here, $V$ is the illuminated volume in the nonlinear medium,
$d(\vec{r})$ is the second-order nonlinearity, and
$\hat{E}_\mu(\vec{r},t)$ ($\mu=p,s,i$) represents the  electric
field operators associated with each of the interacting fields.  In a
nonlinear photonic crystal, each of these fields is described by a
Bloch wave.  Thus, if we assume that the pump field is classical, it
can be expressed as

\begin{equation}\label{Ec:pumpfield}
\hat{E}_{p}^{(+)} \left(\vec{r},t\right)
\rightarrow \int d\omega\quad\alpha_p
\left(\omega\right) E_{K_p}(z,\omega)  \mbox{exp}\left\{i \left[K_p(\omega) z-\omega
t\right]\right\},
\end{equation}
in terms of the Bloch wavenumber $K_p(\omega)$, Bloch envelope
$E_{K_p}(z,\omega)$, and the spectral amplitude $\alpha_p
\left(\omega\right)$.  It is convenient to express the Bloch envelope as a Fourier
series,

\begin{equation}\label{Ec:Bloch_Fourier_pump}
E_{K_p}(z,\omega)=\sum\limits_l \varepsilon_{p l}(\omega) e^{i
G_l z}
\end{equation}
in terms of the spatial harmonics $G_l= 2\pi l/\Lambda$.  The
signal and idler fields are quantized; their positive-frequency
part can be expressed as

\begin{eqnarray}\label{Ec:modefield}
\hat{E}_\mu^{(+)} (\vec{r},t)& = & i  \int d\omega \sum\limits_l \varepsilon_{\mu l}(\omega)
\ell_\mu
(\omega)   \hat{a}_\mu (K_\mu(\omega)+G_l) \nonumber \\
& & \times \mbox{exp}\left(i \left\{[K_\mu (\omega)+G_l] z-\omega
t\right\}\right),
\end{eqnarray}
in terms of the Bloch wave number $K_\mu(\omega)$, envelope
$E_{K_\mu}(z,\omega)$ and the Bloch envelope Fourier series coefficients $\varepsilon_{\mu l}$.  Here, $\ell_\mu (\omega)=\sqrt{\hbar \omega K_\mu'(\omega)/[2
\epsilon_\mu(\omega)S]}$, where $K_\mu'(\omega)$ is the first frequency
derivative of $K_\mu$, $\hat{a}_\mu K(\omega)$ is the annihilation operator
for the signal($s$) or idler($i$) modes,
$\epsilon_\mu(\omega)$ is the permittivity in the nonlinear
medium, and $S$ is the transverse beam area.

It can be shown that the resulting two-photon component of the
state may be written for specific directions of propagation
(throughout this paper assumed to be collinear with the pump
beam) as

\begin{eqnarray}\label{Ec:PDC_state}
|\Psi\rangle&=&\sum\limits_{lmn}\int\int d\omega_s d\omega_i f_{lmn}(\omega_s,\omega_i)\nonumber\\
&&\times a_s^\dag(K(\omega_s)+G_m)a_i^\dag(K(\omega_i)+G_n)
|\mbox{vac}\rangle
\end{eqnarray}
where the joint spectral amplitude $f_{lmn}(\omega_s,\omega_i)$ may be
factored as $f_{lmn}(\omega_s,\omega_i)=\alpha_p(\omega_s+\omega_i)
\phi_{lmn}(\omega_s,\omega_i)$.  Here, $\alpha_p(\omega_s+\omega_i)$
represents the pump spectral envelope function, while
$\phi_{lmn}(\omega_s,\omega_i)$ is a function which describes the
phase-matching properties of the nonlinear photonic crystal and can
be expressed as

\begin{eqnarray}\label{Ec:PMF}
\phi_{lmn}(\omega_s,\omega_i)&=&
\varepsilon_{pl}(\omega_s+\omega_i)\varepsilon_{sm}^*(\omega_s)\varepsilon_{in}^*(\omega_i)
\nonumber\\
&&\times\ell_s(\omega_s)\ell_i(\omega_i)\mbox{sinc}\left( L\Delta
K_{lmn}/2 \right)\nonumber\\
&&\times\exp \left( i L\Delta K_{lmn}/2 \right),
\end{eqnarray}
in terms of the frequency-dependent phase mismatch $\Delta K_{lmn}$
adjusted by the momentum contribution due to the photonic crystal
structure,

\begin{equation}\label{Ec:Pmismatch}
\Delta K_{lmn}=\Delta K+ 2 \pi (l-m-n)/\Lambda
\end{equation}
where $\Delta K=K_p-K_s-K_i$.  The term proportional to $2 \pi/\Lambda$ is similar to that which appears for quasi-phase-matched interactions and has the effect of
shifting the spectral range where phase-matching is attained. Thus,
the phase-matching contributions for different values
of $l-m-n$, if they exist (i.e., if $\Delta k_{lmn}=0$), will tend to be
spectrally distinct from each other.  Note, however, that in general terms, for
photonic crystal periods in the hundreds of nanometers, contributions with
$l-m-n \neq 0$ result in a term proportional to $2 \pi/ \Lambda$
which will tend to be larger than the wave numbers for each of the
three interacting fields, and will therefore also be larger than
$\Delta K$; this makes it difficult to achieve $\Delta K_{lmn} =0$
for $l-m-n \neq 0$.

\begin{figure}[ht]
\centering \epsfg{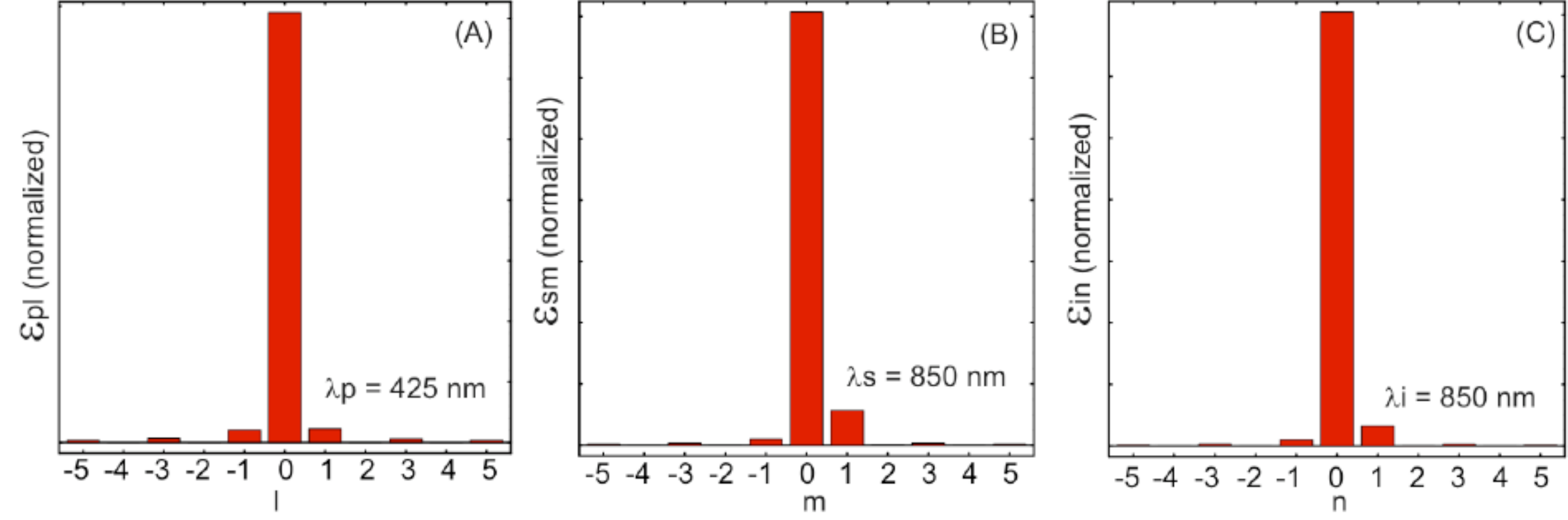}{0.3} \caption{(Color online) Normalized modulus of Fourier coefficients
for (a) an extraordinary wave at $425$ nm, (b) an extraordinary wave
at $850$ nm, and (c) an ordinary wave at $850$ nm.
 \label{Fig:DFourier}}
\end{figure}

Let us now consider the Bloch wave characteristics at $425$ nm and
$850$ nm (which could represent pump and PDC wavelengths) in a
material with the parameters specified at the end of the previous
section. Figure~\ref{Fig:DFourier} shows the modulus of the leading
Fourier series coefficients [see Eqs.(\ref{Ec:Bloch_Fourier_pump})
and (\ref{Ec:modefield})], calculated from the Bloch envelope in
turn determined by the eigenvectors of the translation matrix
which characterizes the periodic material\cite{yariv84}.
Figure~\ref{Fig:DFourier}(a) shows the Fourier coefficients for an
extraordinary wave at $425$ nm, Fig.~\ref{Fig:DFourier}(b) shows
the Fourier coefficients for an extraordinary wave at $850$ nm, and
Fig.~\ref{Fig:DFourier}(c) shows the Fourier coefficients for an
ordinary wave at $850$ nm.  Let us note that a wave at $425$nm
behaves essentially like a plane wave with negligible $m \neq 0$
terms. This is due to the fact for this specific material there is
no band gap in the vicinity of $425$ nm; in particular, the second
order band gap is suppressed by employing a $a/\Lambda=0.5$ duty
cycle. This specific material exhibits a band gap centered at
$904.9$ nm for the ordinary wave and at $876.3$ nm for the
extraordinary wave.  A wave of either polarization at $850$ nm is
sufficiently near to the corresponding band gap so that the
resulting dispersion relation is significantly modified (see
Fig.~\ref{Fig:dispersion}), yet as observed in
Figs.~\ref{Fig:DFourier}(b) and (c small $m=1$ (i.e., reflected
wave) contribution and where other contributions are negligible.

In what follows, we will study the generation of PDC where the emission frequency is in sufficient proximity to the band gap boundaries so that dispersion is strongly modified, though sufficiently removed so that the fraction of light appearing in modes other than the fundamental mode (determined by the values $m,n$) is small; see
Figs.~\ref{Fig:dispersion} and \ref{Fig:DFourier}.  In this regime the signal and
idler photons propagate essentially as plane waves, yet with a
modified dispersion relation with respect to an equivalent source
without a photonic crystal structure.  From
Eq.(\ref{Ec:couple_coef}) there is no second-order band gap for a
0.5 duty cycle, which implies that the effect of the photonic
crystal tends to be insignificant for the pump field, thus
suppressing contributions with $l\neq 0$.  Thus, in this paper, we will concentrate on the contribution $l=m=n=0$, which corresponds to the fundamental, forward-propagating mode for the three interacting fields.

\section{Conditions for factorizability}

In order to carry out an analysis of the relationship between the
various experimental parameters and the resulting spectral
entanglement properties, it is convenient to express the
phase mismatch as a power series in the frequency detunings
$\nu_{s,i}=\omega_{s,i}-\omega_o$, where $\omega_o$ is the
degenerate frequency and $L$ is the crystal length

\begin{eqnarray}\label{Ec:Taylor1}
L\Delta K &\approx& L\Delta K^{(0)}+ \sum\limits_{j=1}^{4}\left( \tau_s^{(j)}\nu_s^j+ \tau_i^{(j)}\nu_i^j \right)\nonumber\\
&&+[2 \tau_p^{(2)}+3 \tau_p^{(3)}(\nu_s+\nu_i)\nonumber\\
&&+2\tau_p^{(4)}(2 \nu_i^2+3\nu_s \nu_i+2 \nu_s^2)]\nu_s
\nu_i+\vartheta(5).
\end{eqnarray}
Here, $\vartheta(5)$ represents fifth, and higher order terms in
the detunings while $\Delta K^{(0)}$ represents the
frequency-independent term, which vanishes when perfect
phase matching occurs at $\omega_o$,

\begin{equation}\label{Ec:PM}
\Delta K^{(0)}=K_s(\omega_o)+K_i(\omega_o)-K_p(2\omega_o).
\end{equation}

Equation(\ref{Ec:Taylor1}) is written in terms of the mismatch in the
$j$th frequency derivative of the wave number between the pump and
the signal and idler wave packets $\tau_{s,i}^{(j)}$, and a term
proportional to the $j$th frequency derivative of the pump
wave number $\tau_{p}^{(j)}$,

\begin{eqnarray}\label{Ec:Taylor_terms}
\tau_\mu^{(j)}&=&L(K_p^{(j)}-K_\mu^{(j)}), \nonumber \\
\tau_p^{(j)}&=&L K_p^{(j)},
\end{eqnarray}
in terms of

\begin{eqnarray}\label{Ec:Taylor_terms2}
K_\mu^{(j)}=(1/j!)(d^j K_\mu/d \omega^j)|_{\omega=\omega_o}, \nonumber\\
K_p^{(j)}=(1/j!)(d^j K_p/d \omega^j)|_{\omega=2\omega_o}.
\end{eqnarray}
%
%
%
%

Let us note that for type-II PDC in standard nonlinear crystals it
is typically sufficient to consider a power series expansion of the
phase mismatch up to the group velocity terms.  However, for NLPCs
with the signal and idler frequencies in  proximity to one of the
band gap boundaries, GVD and higher-order dispersion terms are
strongly enhanced with respect to an equivalent source without a
photonic crystal structure. Thus, it becomes necessary to consider
higher-order terms; here we have considered up to quartic order in
the frequency detunings.

In order to facilitate an analysis of the conditions under which it
is possible to generate two-photon states that are close to
factorable, we write down the joint spectral amplitude in terms of
Gaussian functions.   This can be achieved by modeling the pump
envelope function as a Gaussian function,

\begin{equation}\label{Ec:Pump_function}
\alpha_p(\omega_s+\omega_i)=\exp\left[-(\omega_s+\omega_i-2
\omega_o)^2/\sigma^2\right],
\end{equation}
and by approximating the sinc function in the phasematching
function as a Gaussian funciton i.e. $\mbox{sinc}(x) \approx
\exp(-\gamma x^2 )$ with $\gamma\approx 0.193$.   Under these
approximations, the joint spectral amplitude can be expressed as

\begin{eqnarray}\label{Ec:JSA1}
f(\nu_s,\nu_i)&\approx& \exp \{-(\gamma/4)[L^2 \Delta
K^2+4(\nu_s+\nu_i)^2/(\gamma \sigma^2)] \nonumber\\
&&\ \ \ \ \ \ +i L \Delta K/2 \}.
\end{eqnarray}

By writing $L \Delta K$ in Eq.(\ref{Ec:JSA1}) as its power series
expansion [Eq.(\ref{Ec:Taylor1})], and keeping terms up to fourth
order while assuming that $\Delta K^{(0)}=0$, the joint spectral
amplitude becomes

\begin{equation}\label{Ec:JSA2}
f(\nu_s,\nu_i)\approx\exp\left\{-(\gamma/4)\left[\Phi_s(\nu_s)+\Phi_i(\nu_i)+\Phi_{si}(\nu_s,\nu_i)\right]\right\}
\end{equation}
where

\begin{eqnarray}\label{Ec:JSA3}
&\Phi_\mu(\nu)=[(\tau_\mu^{(1)})^2+4/(\gamma \sigma^2)]\nu^2+2
\tau_\mu^{(1)}\tau_\mu^{(2)} \nu^3\nonumber\\
&+[(\tau_\mu^{(2)})^2+2\tau_\mu^{(1)}\tau_\mu^{(3)}]\nu^4-i
(2/\gamma)\sum\limits_{j=1}^4 \tau_\mu^{(j)}\nu^j
\end{eqnarray}
and

\begin{eqnarray}\label{Ec:exp1}
&\Phi_{si}(\nu_s,\nu_i)/(2 \nu_s
\nu_i)=\tau_s^{(1)}\tau_i^{(1)}+4/(\gamma
\sigma^2)\nonumber\\
&+(2 \tau_p^{(2)} \tau_s^{(1)}+\tau_s^{(2)}\tau_i^{(1)})\nu_s+(2
\tau_p^{(2)} \tau_i^{(1)}+\tau_i^{(2)}\tau_s^{(1)})
\nu_i\nonumber\\
&+(2\tau_p^{(2)} \tau_s^{(2)}+ 3\tau_p^{(3)}
\tau_s^{(1)}+\tau_s^{(3)}
\tau_i^{(1)})\nu_s^2\nonumber\\
&+[2\tau_p^{(2)} \tau_i^{(2)}+ 3\tau_p^{(3)}
\tau_i^{(1)}+\tau_i^{(3)}
\tau_s^{(1)}]\nu_i^2 \nonumber \\
&+[2 (\tau_p^{(2)})^2+\tau_s^{(2)}\tau_i^{(2)}+3\tau
_p^{(3)}(\tau_s^{(1)}+\tau_i^{(1)})] \nu_s \nu_i \nonumber \\
&-i(2/\gamma)[\{ \tau_p^{(2)}+(3/2) \tau_p^{(3)} (\nu_s+\nu_i) +\nonumber \\
&\tau_p^{(4)}(2
\nu_s^2+3 \nu_s \nu_i+2 \nu_i^2)\}]. \nonumber \\
\end{eqnarray}

While $\Phi_{\mu}(\nu)$ (with $\mu=s,i$) represents the
contributions that depend only on one of the two frequencies,
$\Phi_{si}(\nu_s,\nu_i)$ depends on both frequencies, and gives
rise to correlations in the two-photon state;  factorability is
attained if $\Phi_{si}(\nu_s,\nu_i)=0$.  It is apparent from
Eqs.(\ref{Ec:JSA3}) and (\ref{Ec:exp1}) that the dominant terms
which govern the phase matching behavior are the group velocity
mismatch terms, proportional to $\tau_{s,i}^{(1)}$. Let us
consider the effect of making both of these group velocity
mismatch terms vanish, i.e.,

\begin{equation}\label{Ec:GVM}
\tau_s^{(1)}=\tau_i^{(1)}=0.
\end{equation}

In this case, for which the pump, signal, and idler propagate at the
same group velocity,  the expression for the joint spectral amplitude greatly
simplifies, and in particular many of the mixed terms giving rise to
correlations are suppressed. The modulus of the joint spectral
amplitude becomes

\begin{eqnarray}\label{Ec:JSA_approx1}
&|f(\nu_s,\nu_i)| \approx \exp[\{- \nu_s^2 /\sigma^2
-\gamma(\tau_s^{(2)})^2
\nu_s^4/4\}] \nonumber\\
&\times \exp[\{ -\nu_i^2/\sigma^2 -\gamma(\tau_i^{(2)})^2
\nu_i^4/4\}] \nonumber\\
&\times\exp\{-2\nu_s\nu_i/\sigma^2+\gamma\tau_p^{(2)}(\tau_s^{(2)}
\nu_s^3\nu_i+\tau_i^{(2)} \nu_s\nu_i^3)\nonumber\\
&+\gamma[(\tau_p^{(2)})^2+\tau_s^{(2)}\tau_i^{(2)}/2]\nu_s^2\nu_i^2\},
\end{eqnarray}
while the argument becomes

\begin{eqnarray}\label{Ec:argumento}
&\arg[f(\nu_s,\nu_i)]\approx (1/2)\sum\limits_{j=2}^4
(\tau_s^{(j)}\nu_s^j+\tau_i^{(j)}\nu_i^j)+\nu_s\nu_i[\tau_p^{(2)}\nonumber \\
&+3 \tau_p^{(3)} (\nu_s+\nu_i) + 2\tau_p^{(4)}(2 \nu_s^2+3 \nu_s.
\nu_i+2 \nu_i^2)].
\end{eqnarray}

Equation (\ref{Ec:JSA_approx1}) is written as a product of three
exponentials, where the first two represent the factorable
components and the third term describes spectral (modulus)
correlations between the signal and idler photons. Likewise, in
Eq.(\ref{Ec:argumento}) it is the second term, with a $\nu_s \nu_i$
overall multiplicative factor, which describes spectral phase
correlations between the signal and idler photons.

Let us now in addition suppose that the signal and idler photons
experience much stronger dispersion than the pump.  In particular,
let us assume that $j$th-order pump dispersion coefficients
$\tau_p^{(j)}$ may be neglected with respect to the signal and idler
$j$th-order dispersion coefficients $\tau_{s,i}^{(j)}$, i.e.,

\begin{equation}\label{Ec:weak_pump}
|\tau_p^{(j)}|\ll|\tau_s^{(j)}|,|\tau_i^{(j)}|.
\end{equation}

It may be shown that, if the condition in
Eq.(\ref{Ec:weak_pump}) with $j=2$ is satisfied, the fourth order
mixed terms in Eq.(\ref{Ec:JSA_approx1}) proportional to
$\tau_p^{(2)}$ may be neglected.  Similarly, if the condition in
Eq.(\ref{Ec:weak_pump}) with $j=2,3,4$ is satisfied, the mixed
terms in the argument of the joint amplitude
[Eq.(\ref{Ec:argumento})] may be neglected. Imposing this weak pump
dispersion condition, the expression for the joint spectral amplitude is thus further
simplified.  The modulus may now be written as

\begin{eqnarray}\label{Ec:JSA_approx2}
|f(\nu_s,\nu_i)| &\approx& \exp[-(\nu_s+\nu_i)^2/\sigma^2\nonumber\\
&&-(\gamma/4)(\tau_s^{(2)} \nu_s^2+\tau_i^{(2)} \nu_i^2)^2]
\end{eqnarray}
while the argument may be written as

\begin{equation}\label{Ec:argumento2}
\arg[f(\nu_s,\nu_i)]\approx (1/2)\sum\limits_{j=2}^4
(\tau_s^{(j)}\nu_s^j+\tau_i^{(j)}\nu_i^j]).
\end{equation}

Let us note that while the modulus contains mixed terms
proportional to $\nu_s \nu_i$ and to $\nu_s^2 \nu_i^2$, the
argument does not contain mixed terms.  Let us now assume, in
addition to group velocity matching [see Eq.(\ref{Ec:GVM})] and
weak pump dispersion (see Eq.(\ref{Ec:weak_pump})] that the pump is
broadband.  It may be shown from Eq.(\ref{Ec:JSA_approx2}) that the
joint spectral amplitude exhibits no dependence on the pump
bandwidth if

\begin{equation}\label{Ec:pump}
\sigma \gg 2(4/\gamma)^{1/4}(\tau_s^{(2)}+\tau_i^{(2)})^{-1/2}.
\end{equation}

Thus, for a sufficiently broadband pump [so that Eq.(\ref{Ec:pump})
is satisfied], the modulus of the joint spectral amplitude reduces
to

\begin{equation}\label{Ec:JSA_approx3}
|f(\nu_s,\nu_i)| \approx \exp[-(\gamma/4)(\tau_s^{(2)}
\nu_s^2+\tau_i^{(2)} \nu_i^2)^2].
\end{equation}

We have seen that when the following three conditions are satisfied:  (1) complete group velocity matching [see Eq.(\ref{Ec:GVM})], (2) weak pump dispersion [see Eq.(\ref{Ec:weak_pump})], and (3) sufficiently broadband pump [see Eq.(\ref{Ec:pump})], the joint spectral amplitude attains a particularly simple form in
which there is a single mixed term, proportional to $\nu_s^2
\nu_i^2$ up to fourth order in $\Delta k^2$.   For the specific cases we analyzed (see the next section), while conditions 1 and 3 must be satisfied in order to attain nearly factorable two-photon states, the effect of the mixed terms controlled by condition 2 is comparatively small.  In the next section, we will show that the properties of one-dimensional nonlinear photonic crystals, may be exploited for the fulfilment of conditions 1 and 3, and partial fulfilment of condition 2.

\section{Fulfilment of conditions with nonlinear photonic crystals}

Our strategy for controlling the spectral properties of PDC photon
pairs generated by one-dimensional NLPCs is based on the
observation that the group velocity is sharply reduced for
frequencies in the vicinity of the band gap boundaries, as
illustrated in Fig.~\ref{Fig:dispersion} (b).    It is possible to
exploit the properties of photonic crystals to compensate for
material dispersion and to impose specific conditions on the group
velocities of the three interacting fields. Indeed, while for a
standard optical material the pump will tend to propagate at a
lower group velocity than the generated light, it is possible to
design a NLPC so that the group velocities of the signal and idler
photons in  proximity to one of the band gap boundaries,  are
reduced sufficiently to make them equal to the pump group
velocity.  The former can be achieved while maintaining
essentially a plane-wave character  for the corresponding signal
and idler Bloch waves.  Likewise, the NLPC can be designed so that
the pump frequency is far from band gaps, so that the pump field is
essentially unaffected by the photonic crystal structure. Indeed,
NLPCs make it possible to attain, even in a type-II geometry
where each of the three fields experiences different dispersion
characteristics, complete group velocity matching where
$K_p'=K_s'=K_i'$, or in the notation of the previous section
$\tau_s^{(1)}=\tau_i^{(1)}=0$.

In order to simplify our analysis of realistic NLPCs, we first
limit the parameter space; we assume that the crystal is operated
at room temperature and assume a duty cycle $a/\Lambda=0.5$
throughout. Likewise, we regard the degenerate PDC frequency
$\omega_o$ as a fixed parameter.  It is indeed remarkable that,
in general, nearly factorable PDC photon pair sources based on NLPCs permit the
specification of an arbitrary central emission wavelength, and for
that matter an arbitrary material, constrained only by the ability
to fabricate the crystal with the required periodic index of refraction variation.
Thus, for degenerate collinear PDC at a given freely specified
central frequency, we are left with the following experimental
parameters: Bragg period ($\Lambda$), permittivity contrast
($\alpha$), crystal propagation angle ($\theta_{pm}$), crystal
length ($L$), and pump bandwidth ($\sigma$).

In what follows we present a numerical analysis for the simultaneous
fulfilment of basic phase matching [$\Delta K^{(0)}=0$; see
Eq.(\ref{Ec:PM})] and complete group velocity matching [see
Eq.(\ref{Ec:GVM})] for a one-dimensional NLPC based on a BBO crystal.
Let us note that these conditions are independent of crystal length
and pump bandwidth. Thus, we have three variables: Bragg period
($\Lambda$), permittivity contrast ($\alpha$), and crystal
propagation angle ($\theta_{pm}$) with which to satisfy three
conditions: (i) $\Delta K^{(0)}=0$ (ii) $K_p'=K_s'$, and (iii)
$K_p'=K_i'$ (where the pump frequency derivatives are evaluated at
$2 \omega_o$ while signal and idler frequency derivatives are evaluated
at $\omega_o$).

\begin{figure}[ht]
\centering \epsfg{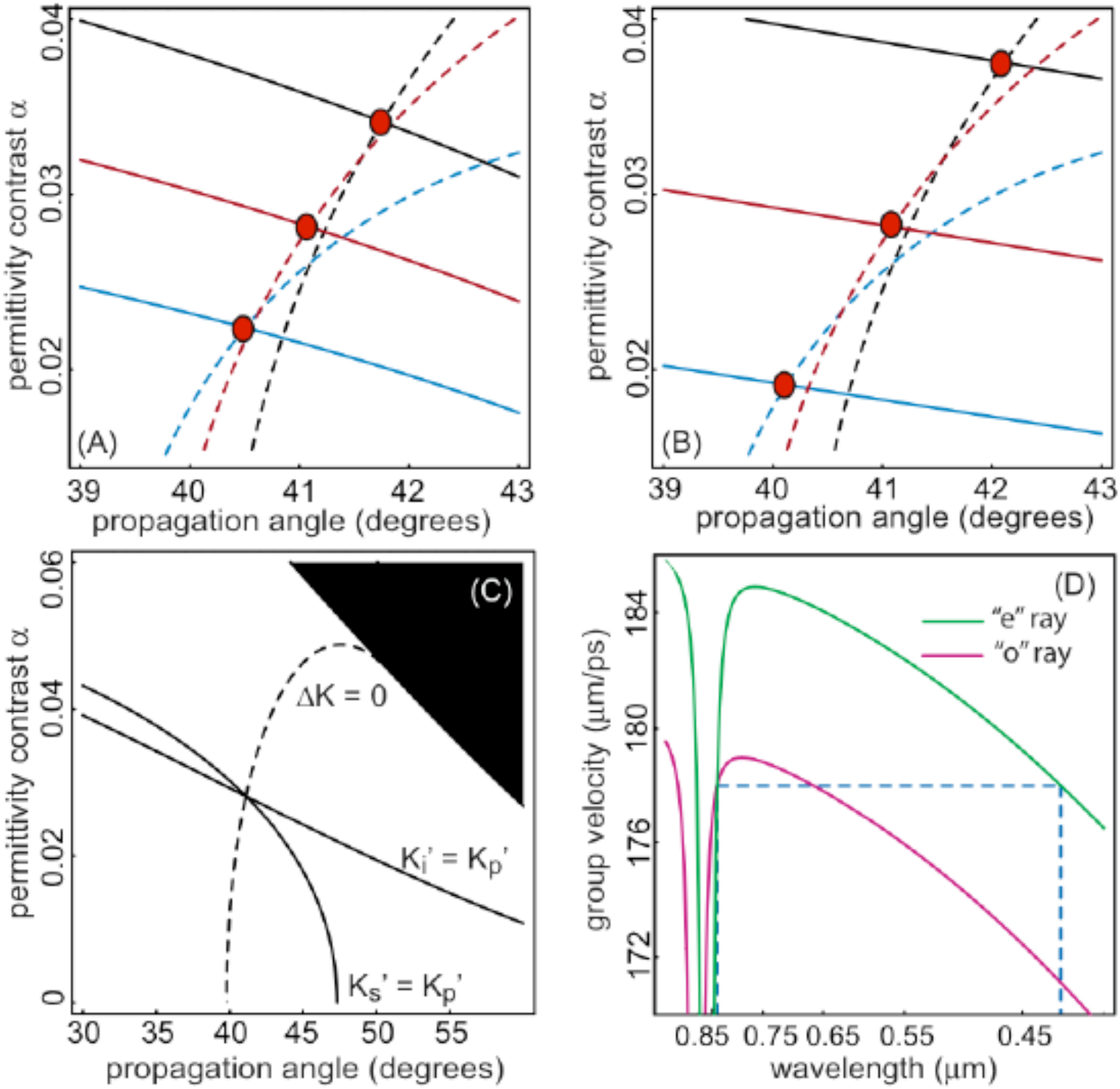}{0.53} \caption{(Color
online) (a) Graphical representation of the conditions (i) $\Delta
K=0$ and (ii) $K_p'=K_s'$, and (b) the conditions (i) $\Delta K=0$
and (ii) $K_p'=K_i'$ on $\{\alpha,\theta_{pm}\}$ space, for
constant $\Lambda$  [we have used the values $\Lambda=277.8$ nm
(shown in blue), $\Lambda=274.9$ nm (red) and $\Lambda=271.9$nm
(black)]. Each interesection point indicates a solution for
simultaneous phasematching and group velocity matching between the
pump and one of the generated photons. (c) Graphical
representation of the conditions (i) $\Delta K=0$, (ii) $K_p'=K_s'$
y (iii) $K_p'=K_i'$ on $\{\alpha,\theta_{pm}\}$ space, assuming
$\Lambda=274.9$nm.  The interesection point indicates a solution
for simultaneous phase matching and \textit{complete} group
velocity matching. Note that, in the shaded area, the wave number
becomes complex, and therefore light does not propagate. (d) Group
velocity vs frequency for the ordinary and extraordinary waves;
the dotted (black) lines indicate the resulting complete group
velocity matching.
 \label{Fig:vg_empcompleto}}
\end{figure}

Figure~\ref{Fig:vg_empcompleto}(a) shows in $\{\alpha,\theta_{pm}\}$
space, for three different values of $\Lambda$, the contour defined by the
condition $\Delta K^{(0)}=0$ together with the contour defined by
the condition $K_p'=K_s'$. Note that both contours shift as the
Bragg period $\Lambda$ is modified. Similarly,
Fig.~\ref{Fig:vg_empcompleto}(b) shows, for the same three values of $\Lambda$,
the contour defined by the condition $\Delta K^{(0)}=0$ together
with the contour defined by the condition $K_p'=K_i'$. There
exists a specific value of the Bragg period $\Lambda$ for which
the three contours meet a single point on $\{\alpha,\theta_{pm}\}$
space, yielding the specific values of the three parameters
($\alpha$, $\theta_{pm}$, and $\Lambda$) which satisfy
simultaneously phase matching and complete group velocity matching.
Assuming $\lambda_{o}=2 \pi c/\omega_0=850$ nm (which we stress
may be freely specified), these values are $\alpha=0.028$,
$\theta_{pm}=41.1^\circ$, and $\Lambda=274.9$ nm, obtained numerically
from the intersection of the three resulting contours [see
Fig.~\ref{Fig:vg_empcompleto}(c)]. For this choice of parameters,
Fig.~\ref{Fig:vg_empcompleto}(D) shows a plot of the group
velocities for the ordinary and extraordinary waves.  In this plot
we indicate that, for a type-II interaction where the pump is an
extraordinary wave and the signal and idler are extraordinary and
ordinary, respectively, we indeed obtain identical group velocities
for the three fields.  Let us note that to our knowledge no other
reported technique permits complete group velocity matching in a
collinear, type-II nonlinear parametric interaction.

According to the conditions derived in the previous section, in
order to guarantee factorability we require the fulfilment of the
weak dispersion condition [see Eq.(\ref{Ec:weak_pump})], in
addition to complete group velocity matching.  In particular, if
this condition is satisfied for $j=2$, then three of the four
remaining mixed terms in the joint spectral amplitude (to fourth
order in the detunings) may be suppressed.   As illustrated by
Fig.~\ref{Fig:dispersion}, when the signal and idler frequencies
are in proximity to one of the band gap boundaries while the pump
experiences essentially only the material dispersion, group
velocity dispersion and higher order terms associated with the
generated light will be greatly enhanced with respect to
corresponding pump quantities. Thus, for the specific experimental
parameters yielding complete group velocity matching (see the previous
paragraph), the resulting second order dispersion coefficients
are $\tau_s^{(2)}=6.12 $fs$^2$, $\tau_i^{(2)}=0.87 $fs$^2$ and
$\tau_p^{(2)}=0.16 $fs$^2$. The condition
$|\tau_p^{(2)}|\ll|\tau_s^{(2)}|,|\tau_i^{(2)}|$ will be more
accurately satisfied the closer the degenerate PDC frequency is to
the band gap boundaries for each polarization.  However, note that,
because the band gap boundaries for the two polarizations are
spectrally distinct, the weak pump dispersion condition cannot be
satisfied to the same degree for both photons.  Thus, for this
specific choice of parameters, while the three mixed terms
proportional to $\tau_p^{(2)}$ are reduced, they are not
suppressed perfectly.

\begin{figure}[ht]
\centering \epsfg{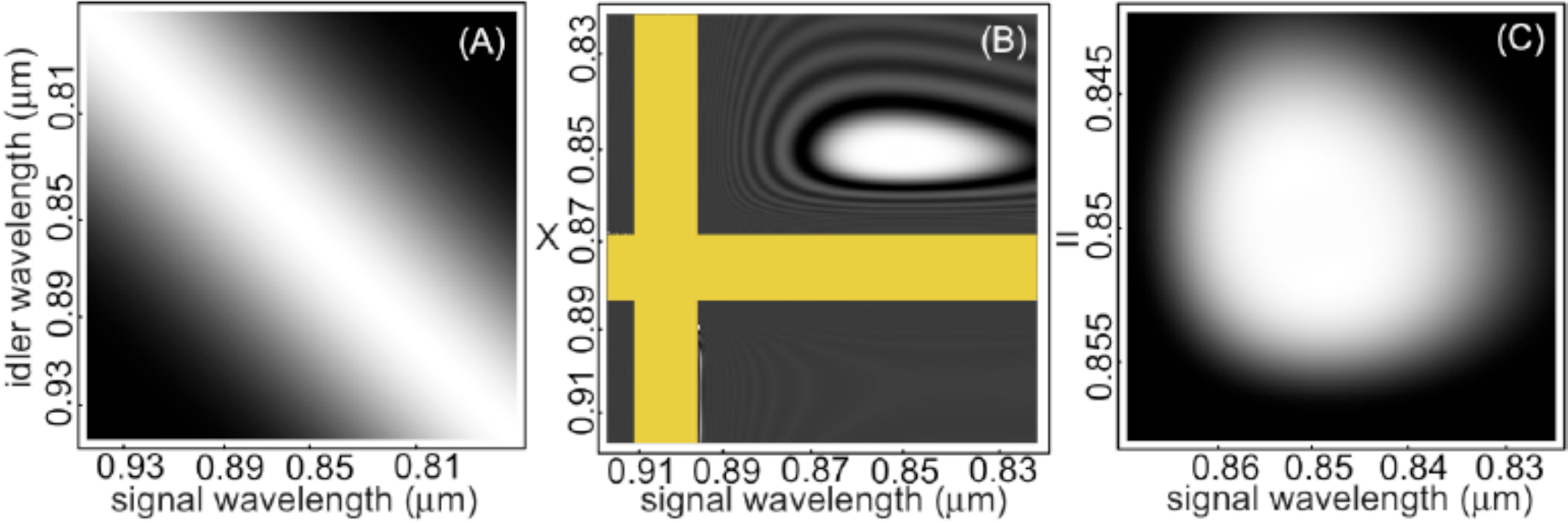}{0.39} \caption{(Color online) Joint spectral
intensity, where dispersive effects to all orders are included and
where the Gaussian approximation has not been used.  (a) Pump
spectral envelope with FWHM bandwidth of $10$ nm. (b)
Phase matching function; the two yellow bands (light gray) indicate the band gaps for each of the two polarizations. (c) Joint spectral intensity, given as the product of the pump envelope and phase matching function.
 \label{Fig:estado_completo}}
\end{figure}

The third condition for factorability derived in the previous
section, apart from complete group velocity matching and weak pump
dispersion, is that the pump should be sufficiently broadband [see
Eq.(\ref{Ec:JSA3})].  Note that while the Bragg period, permittivity
contrast and crystal propagation angle must have specific values in
order to guarantee complete group velocity matching, the pump
bandwidth and crystal length must fulfil a comparatively soft
condition given as the inequality in Eq.(\ref{Ec:pump}).
Figure~\ref{Fig:estado_completo} represents the resulting joint
spectral intensity $|\alpha_p(\omega_s+\omega_i)\phi_{000}(\omega_s,\omega_i)|^2$, for a pump centered
at $\lambda_p=2\pi c/(2\omega_o)=425$ nm with a full width at half
maximum bandwidth of $10$ nm, with crystal length $L=4$ mm and where
the rest of the parameters are as specified above, in the context of
Fig.~\ref{Fig:vg_empcompleto}. Note that in this plot we have taken
into account the complete dispersion (rather than a truncated power
series approach) and have not used the Gaussian approximation. While
these approximations were essential for our analysis, it is
important to verify the degree of factorability in a two-photon
state produced by a realistic source.
Figure~\ref{Fig:estado_completo}(a) shows the pump spectral envelope
$\alpha_p(\omega_s+\omega_i)$ as given by
Eq.(\ref{Ec:Pump_function}), Fig.~\ref{Fig:estado_completo}(b) shows
the phasematching function $\phi_{000}(\omega_s,\omega_i)$ as given
by Eq.(\ref{Ec:PMF}), and Fig.~\ref{Fig:estado_completo}(c) shows the
joint spectral intensity. It
is apparent from the figure that the resulting two-photon state is
nearly factorable.  Indeed, a numerical Schmidt decomposition~\cite{law00}
yields a Schmidt number of $K \approx 1.11$ (which could be reduced further by moderate spectral filtering).

As has already been discussed, the effective control over group velocity properties permitted by photonic crystal structures, which may be exploited to obtain nearly factorable two-photon states, enhances higher-order dispersion terms.  This has the effect that, when complete group velocity matching is not fulfilled, the contours of equal amplitude of the phase matching function in {$\omega_s,\omega_i$} space tend to be highly curved.  Therefore, the technique presented here is not naturally suited for the generation of states exhibiting strict correlation or anticorrelation in frequency, unless the bandwidth of interest is small enough that the curvature of the phase matching function may be neglected.

\begin{figure}[ht]
\centering \epsfg{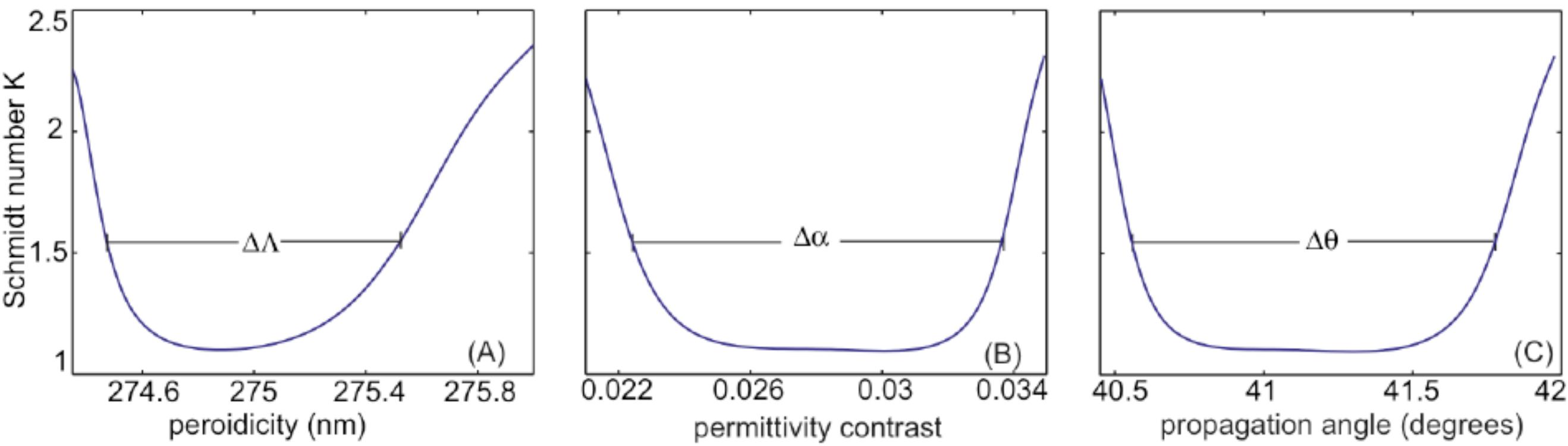}{0.32} \caption{(Color online) Schmidt number $K$
plotted vs (a) Bragg period $\Lambda$, (b) permittivity contrast
$\alpha$, and (c) crystal propagation angle $\theta_{pm}$.
 \label{Fig:tolerances}}
\end{figure}

For practical implementations of this technique, it is important
to analyze the required tolerances for the experimental
parameters, in order to ensure factorizability.
Figure~\ref{Fig:tolerances} shows a plot of the Schmidt number $K$
as a function of the three parameters which must satisfy a strict
condition in order to attain complete group velocity matching. We
plot the Schmidt number as a function of each of these parameters,
while maintaining all others equal to their nominal values (which
yield the minimum value of $K$).  Thus,
Fig.~\ref{Fig:tolerances}(a) shows $K$ vs the Bragg period
$\Lambda$, Fig.~\ref{Fig:tolerances}(b) shows $K$ vs the
permittivity contrast $\alpha$, and Fig.~\ref{Fig:tolerances}(c)
shows $K$ vs the crystal propagation angle $\theta_{pm}$.   We
define the tolerance in variable $x$ as $\Delta x= x_2-x_1$  where
$x_{1,2}$ (with $x_2>x_1$) are the values where $K$ rises to
$K=\sqrt{2} K_{min}$ in terms of the value of the Schmidt number
attained when all variables are equal to their nominal values
$K_{min}$.  Thus, we obtain $\Delta \Lambda\approx 1.04$ nm,
$\Delta \alpha\approx 0.011$, and $\Delta \theta_{pm}\approx
1.23^\circ$.

\section{Conclusions}

We have studied the generation of photon pairs by spontaneous
parametric down-conversion in a one-dimensional nonlinear photonic
crystal, with a broadband pump.  We have considered a NLPC based
on a standard $\chi^{(2)}$ crystal with uniaxial birefringence
where the dispersion for both polarizations has been altered from
its natural state so that, throughout the length of the resulting
crystal, the index of refraction for each polarization alternately
takes the value consistent with material-only dispersion and a
value that is larger than the former, consistent with a small
permittivity contrast.  We have developed a set of conditions
that must be satisfied in order to guarantee a nearly factorable
two photon state. These conditions are: (i) complete group velocity
matching, where the pump pulse and the signal and idler photons
propagate at the same group velocity [see Eq.(\ref{Ec:GVM})], (ii)
weak pump dispersion [characterized by coefficients
$\tau_p^{(j)}$, see Eq.(\ref{Ec:Taylor_terms})], relative to
signal and idler dispersion [characterized by coefficients
$\tau_{s,i}^{(j)}$, see Eq.(\ref{Ec:weak_pump})], and (iii)
sufficient pump bandwidth [see Eq.(\ref{Ec:pump})].  We have shown
that the strongly-modified dispersion in the spectral vicinity of
the bandgap boundaries in a NLPC may be exploited in order to
fulfil condition (i), and to partially fulfil condition (ii).  The
essential advantage of this technique is that for an arbitrary
nonlinear material operated at an arbitrary spectral range it is
in principle possible to design a photonic-crystal structure which
counteracts the material dispersion so as to permit the generation
of factorable photon pairs in a type-II degenerate, collinear
geometry which is compatible with wave guiding and which permits
high efficiency photon pair splitting.  It is, however,
anticipated that the fabrication of the appropriate
photonic-crystal structure may pose the main technical challenge
in practical implementations.  Longer Bragg periods, potentially
more easily fabricated, could be exploited through higher-order
bandgaps. It is expected that this work may be important for the
development of optimized nonclassical light sources for quantum
information processing.

\begin{acknowledgements}
This work was supported by Conacyt grant No. 46370-F and UC-MEXUS grant No.
CN-06-82.
\end{acknowledgements}


\end{document}